\documentclass[oneside]{article}
\usepackage[T1]{fontenc}
\usepackage{authblk}
\usepackage{float}
\usepackage{amsmath}
\usepackage{graphicx}
\usepackage{xfrac}
\usepackage[english]{babel}
\usepackage{lmodern}
\usepackage[T1]{fontenc}
\usepackage[latin9]{inputenc}
\usepackage{mathtools}
\usepackage{graphicx}
\usepackage{esint}
\usepackage[unicode=true,
 bookmarks=false,
breaklinks=false,pdfborder={0 0 1},backref=false,colorlinks=false]
 {hyperref}
\usepackage[a4paper, total={210mm,297mm},margin=2.0cm]{geometry}

\usepackage{datetime}
\newdate{date}{2}{12}{2019}

\title{Lattice Boltzmann simulations capture the multiscale physics of soft flowing crystals}

\author[1]{A. Montessori}

\author[1,2]{A. Tiribocchi}

\author[1,2]{F. Bonaccorso}

\author[1]{M. Lauricella\thanks{Electronic address: \texttt{m.lauricella@iac.cnr.it}; Corresponding author}}

\author[1,2,3]{S. Succi}

\affil[1]{Istituto per le Applicazioni del Calcolo CNR, via dei Taurini 19, Rome, Italy}
\affil[2]{Center for Life Nano Science@La Sapienza, Istituto Italiano di Tecnologia, 00161 Roma, Italy}
\affil[3]{Institute for Applied Computational Science, John A. Paulson School of Engineering and Applied Sciences, Harvard University, Cambridge, USA}

\date{\displaydate{date}}

\begin{document}

\maketitle

%%%% Abstract text to be placed here %%%%%%%%%%%%
\begin{abstract}
The study of the underlying physics of soft flowing materials depends heavily on numerical simulations,
due to the complex structure of the governing equations reflecting the competition of concurrent mechanisms
acting at widely disparate scales in space and time. A full-scale computational
modelling remains a formidable challenge since it amounts to simultaneously handle
six or more spatial decades in space and twice as many in time. Coarse-grained methods often provide a viable strategy to significantly mitigate this issue,
through the implementation of mesoscale supramolecular forces designed to capture the essential physics at a fraction of the computational cost of a full-detail description.

Here, we review some recent advances in the design of a lattice Boltzmann mesoscale approach
for soft flowing materials, inclusive of near-contact interactions (NCI) between dynamic interfaces,
as they occur in high packing-fraction soft flowing crystals.

The method proves capable of capturing several aspects of the rheology of soft flowing crystals,
namely, i) a $3/2$ power-law dependence of the dispersed phase flow rate on the applied pressure gradient,
ii) the structural transition between an ex-two and ex-one (bamboo) configurations with the associated drop of the flow rate,
iii) the onset of interfacial waves once NCI is sufficiently intense.
\end{abstract}
%%%%%%%%%%%%%%%%%%%%%%%%%%%

%%%%%%%%%% Insert the texts which can accomdate on firstpage in the tag "fmtext" %%%%%

\section{Introduction}

In the last few decades, the condensed matter has moved great lengths in the direction of soft matter, namely the study of complex states of matter at the interface between the three basic ones: gas, liquid and solid \cite{fernandez2016,piazza2012,lavren2003}. 
Besides their practical relevance for a host of applications in chemistry, material science and biology, soft materials also raise a fundamental challenge to non-equilibrium thermodynamics, since they display properties which cannot be traced back to any of the three fundamental states they derive from \cite{jones2002,degennes1994}. 
For instance, foams, binary mixtures of gas and liquid, each featuring linear Newtonian rheology,
exhibit highly non-Newtonian mechanical and rheological properties.

This has stimulated an outburst of experimental and theoretical activity alike, including computational methods, aiming at shedding light into the underlying mechanisms controlling the behaviour of soft matter systems \cite{bernaschi2019,schiller2018,henrich2018}.

Computational methods can often illuminate regions of parameter space not accessible to experiments and explore non-perturbative regimes totally beyond the reach of analytical methods \cite{prosperetti2007,kartunnen2009,kruger2013numerical}. 
Yet, they face with a formidable multiscale challenge, since many soft matter systems host concurrent 
interactions encompassing six or more decades in space (from tens of $nm$ to $cm$) and easily twice 
as many in time (from $ps$ to $s$) \cite{bernaschi2019}.

Under such state of affairs, two major avenues open up: the first consists in developing highly sophisticated multiscale methods capable of covering five-six spatial decades through a clever combination of advanced computational techniques, such as local grid-refinement, adaptive grids, or grid-particle combinations \cite{fili98,stein2017,dupuis2007,distaso2016,bernaschi2009}. 
Such methods are potentially compelling but extremely demanding, both in terms of programming 
and day-by-day operational complexity.

The second avenue consists in developing suitable coarse-grained models, operating at the mesoscale, say microns, through the incorporation of effective forces and potentials designed in such a way as to retain the essential effects of the fine-grain scales on the coarse-grained ones \cite{succi2018lattice,yeomans2006}. 
While inherently approximate in nature, and strongly dependent on the degree of universality offered by the 
specific problems at hand, the latter alternative is very appealing because, when it works, it comes with 
game-changing computational savings.

In this article, we shall present a series of results within the framework of the latter strategy. In particular, we shall discuss the application of a new class of mesoscale lattice Boltzmann models for multi-component fluids, accounting for near-contact interactions, occurring at the molecular scale around tens of nanometers \cite{montessori2019jfm,montessori2019mesoscale}.

The effectiveness of this approach will be showcased for a number of problems pertaining to soft flowing crystals, namely ordered collections of flowing droplets, generated by microfluidic devices.  In each and every case discussed in this paper, we highlight 
the specific contribution of the simulations to a better {\it understanding} of the basic phenomenon in point, both 
in terms of interpreting existing data and also in terms of proposing new experiments and devices.

The organization of this article is as follows. Section \ref{R0} provides an overview of recent experimental results of soft flowing crystals. In Section \ref{R1}, we illustrate the process of droplet generation while in  Sections \ref{R2} and \ref{R3} we elucidate the role of the velocity field in shaping up different soft crystal 
configurations.
The details of the simulation approach are reported in Materials and Methods.

\section{Microfluidic soft flowing crystals}\label{R0}

Microfluidic crystals consist of a highly ordered and uniform mesoscale porous matrix, made by a dispersion of a dense emulsion of droplets (air bubbles or fluid droplets, such as water) stabilized by a surfactant, embedded in a continuous phase, such as oil \cite{raven2009microfluidic,marmottant2009microfluidics,vecchiolla2019dislocation,fei2020discrete}. This soft material is usually produced within microchannels, structures a few dozens micrometres in width and manufactured by soft-lithography techniques \cite{duffy1998}. A famous example in point is offered by the flow focusers \cite{anna2007,garstecki2004,garstecki2006flowing} (see Fig.~\ref{fig1}), microdevices currently primarily employed for the production of mono-dispersed microemulsions, due to the capability of handling very tiny amounts of fluids (of the order of the picolitre \cite{squires2005}) and to the accurate control over monodispersity and droplet size \cite{marmottant2009microfluidics,dolletprl}.

Such devices are made of three channels supplying the dispersed phase (from the horizontal branch) and the continuous phase (from the two vertical branches) through a tiny orifice, located downstream of the three coaxial inlet flows (see Fig.\ref{fig1}). The three inlet channels (the two vertical ones and the horizontal one) have a width of $H\simeq 200\mu$m, the striction is approximately $h\simeq 100\mu$m, and the outlet channel width is $H_c\simeq 100\mu$m. The height of the device is $W\simeq 100\mu$m. Since in these systems capillary forces dominate over viscous ones, the capillary number is usually rather small. By assuming a dynamic viscosity of the dispersed phase (water) $\mu\simeq 10^{-3}$$Pa\cdot s$, its inlet velocity $u_d\sim 0.1$$m/s$ and an interface tension of the oil-water mixture approximately $50$$mN/m$, one gets $Ca\sim 10^{-3}$, a value in very good match with those characterizing  flow focuser devices~\cite{marmottant2009microfluidics}. 

The mechanism of droplet formation relies on the periodic pinch-off of the dispersed jet by the continuous stream occurring in the orifice \cite{garstecki2005,marmottant2009microfluidics,montessoriprf}. In particular, the droplet size is controlled by the pressure of the dispersed phase and by the flow rate of the continuous phase. The high degree of reproducibility is guaranteed by the absence of hydrodynamic instabilities since these devices operate in a regime where viscous forces largely dominate over inertia (i.e. the Reynolds number of the flow is much smaller than 1) \cite{marmottant2009microfluidics,garsteckiprl2005}. This stabilizes the pinch-off process and produces highly precise droplet sizes. When the density of the droplets in the outlet channel is sufficiently high, a flowing crystal can be produced, typically formed by an ordered array of self-assembled bubbles/droplets exhibiting a hexagonal crystal-like pattern (see Fig.\ref{fig2}) \cite{marmottant2009microfluidics,raven2009microfluidic}.   

Besides being of interest {\it per se} as an intriguing example of complex state of flowing matter, soft flowing crystals may find good use in a wide class of advanced technological applications. Typical examples are catalyst support in porous materials, electrochemical sensing, and template for tissue engineering, where the trabecular morphology is fundamental for its correct functioning \cite{kimmins2010,costantini2014,robinson2014}. 

A pre-patterned design of a soft flowing crystal crucially relies on a careful control over multiple parameters, such as surface tension and viscosity of the mixture, inlet pressure of the dispersed phase and flow rate of the continuous one, dimensions of orifice and channels and, ultimately, near-contact forces (such as nanoscopic van der Waals and disjoining pressure, to name but a few) occurring in the thin film of fluid separating droplet interfaces. Despite their short-range nature, the effect of such forces may encompass several lengthscales, from the nanometer up to the sub-millimetre size, leading to long-range rearrangements of the emulsion structure in the outlet channel.  In the worst scenario, such rearrangements could disrupt the soft crystal, thus compromising the final material design. 

In the following, we show that many morphological structures, as well as dynamic features of the soft flowing crystals observed at the sub-micrometre scale, can be remarkably well captured using an innovative lattice Boltzmann approach which relies exclusively on a mesoscale description of the underlying physics \cite{montessori2019jfm}. In particular, the model (i) correctly reproduces the process of droplet generation within the microfluidic device, (ii) fully describes the complex non-equilibrium transition regimes between different flowing crystal configurations and (iii) recovers the dynamics of near-contact deformable interfaces. Here we illustrate the points (i) and (ii), while we remind to Ref.\cite{tiribocchipre2019} for an extensive discussion of the interface fluctuations in thin films of fluid.

These results support the view that, at least for this system, the present lattice Boltzmann method offers a unique numerical platform providing a realistic description of the physics across almost five lengthscales, through a minimal set of control parameters directly accessible to experiments.

\section{Results}

\subsection{Droplet generation in microfluidic devices}\label{R1}

Marmottant et al. \cite{raven2009microfluidic,marmottant2009microfluidics} have shown that a soft flowing microfluidic crystal can be designed by dispersing air bubbles in fluid phase within a flow focuser. The primary mechanism causing droplet formation relies on a combined effect of the pressure drop, due to sudden channel expansion, and of the shear stress exerted by the continuous phase inside the nozzle, which squeezes the dispersed phase inwards and causes its rupture \cite {montessoriprf,raven2009microfluidic,marmottant2009microfluidics}. When this occurs, the generated droplet is pushed forward by a novel droplet produced at the rear, in a recurrent process in which the resulting emulsion moves almost unidirectionally in the outlet channel. Importantly, droplet formation is controlled by the pressure of the gas set at the inlet channel.

In our numerical experiment, we consider a soft flowing crystal made of two immiscible fluids, such as water drops (the dispersed phase) embedded in oil (the continuous phase). Despite the different nature of the dispersed phase, an almost identical mechanism for the production of microfluidic crystals is found to hold.

Here, droplet formation and structure of the resulting emulsion can be controlled by tuning the dispersed-to-continuous flow rate ratio, defined as $\phi=u_d/2u_c$ (where $u_d$ and $u_c$ are the speeds of the dispersed and continuous phases, respectively, at the inlet channel). Further, but no less remarkable, the droplet formation is also driven by the strength of the near-contact forces $F_{rep}$, a mesoscopic force field modelling the repulsive effect of a surfactant confined at the fluid interface and preventing droplet coalescence (see Materials and Methods for the details of its numerical implementation).

However, at the interface lengthscale, capillary effects, controlled by the surface tension, can favour droplet merging. 

A suitable dimensionless number accounting for the competition between surface tension $\sigma$ and near-contact forces $F_{rep}$ 
can be defined as  ${\cal N}_c=A\Delta x/\sigma$, where $A$ sets the magnitude of the repulsive force and $\Delta x$ is 
the minimum distance of interaction between two interfaces in close contact, usually ranging from $10$ to $100$ $nm$ \cite{derjaguin1940} 
(see Fig.~\ref{fig5} in Materials and Methods). Typically, if ${\cal N}_c \ll 1$, capillary effects dominate and droplets coalesce, 
whereas at ${\cal N}_c\sim 1$ near-contact interactions prevail and droplet merging is inhibited.

\begin{figure}
\includegraphics[scale=0.5]{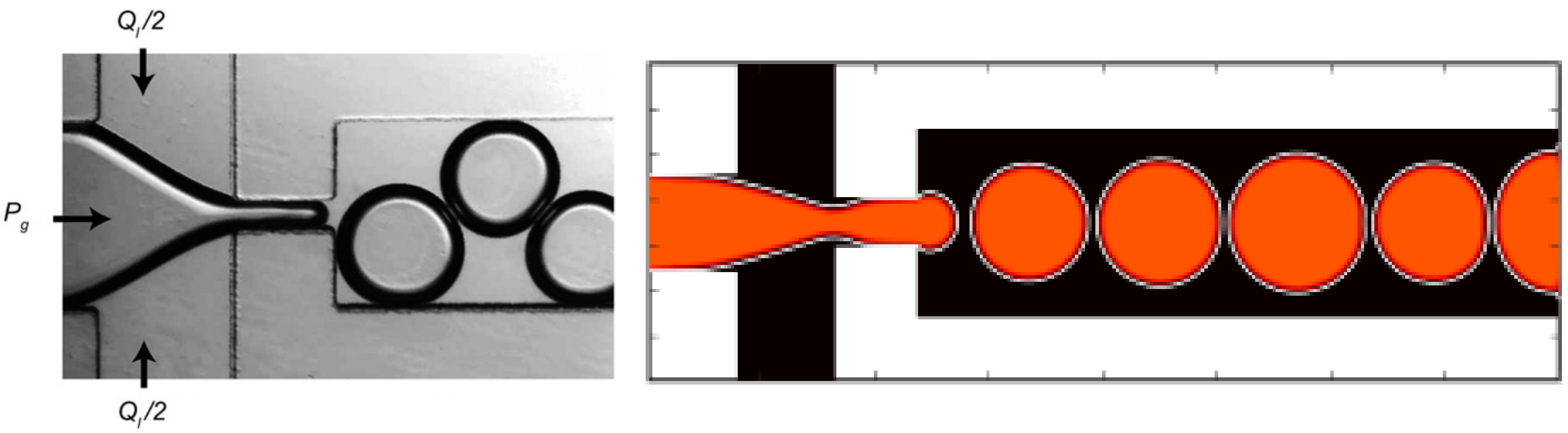}
\caption{\label{fig1} Left: Microfluidic flow focuser used for the production of air bubbles in Ref.\cite{raven2009microfluidic}. A gas, injected with pressure $P_g$ from the horizontal branch, and water, ingjected with  flow rate $Q_l/2$ from the vertical branches, are focussed into the orifice of width=$100$$\mu$m. Air bubbles are then collected in an outlet chamber located downstream. Right: Simulation of production of a microfluidic crystal made of fluid droplets (red)  dispersed in an immiscible secondary fluid (black).}
\end{figure}

A typical arrangement is sketched in Fig.\ref{fig1} (right) obtained for $\phi=1$ and ${\cal N}_c=0.1$. 
The dispersed phase (in red in Fig.\ref{fig1}) is produced at a predefined speed $u_d$ in the horizontal branch of the inlet channel, while the continuous phase (in black) comes from the two vertical branches at speed $u_c$.  
They are both driven to the striction where the break-up process occurs \cite{montessoriprf}, and, finally, a droplet emerges in the outlet channel.

Even though the configurations shown in Fig.\ref{fig1} are still far from a flowing microfluidic crystal, it is remarkable that, by solely controlling the two independent parameters $\phi$ and ${\cal N}_c$, droplet generation can be simulated to a 
high degree of accuracy by our lattice Boltzmann approach. 

Such results invite the following questions: Can this model capture more complex flowing crystal patterns and 
particularly the non-equilibrium structural transitions, such as those reported 
in recent works by Marmottant~\cite{raven2009microfluidic,marmottant2009microfluidics}? Also, to which extent does this minimal mesoscale description provide results readily comparable with experiments?

In the next section, we focus precisely on these two aspects. 

\subsection{Non-equilibrium transition between different flowing crystal patterns}\label{R2}

\begin{figure}
\includegraphics[scale=1.0]{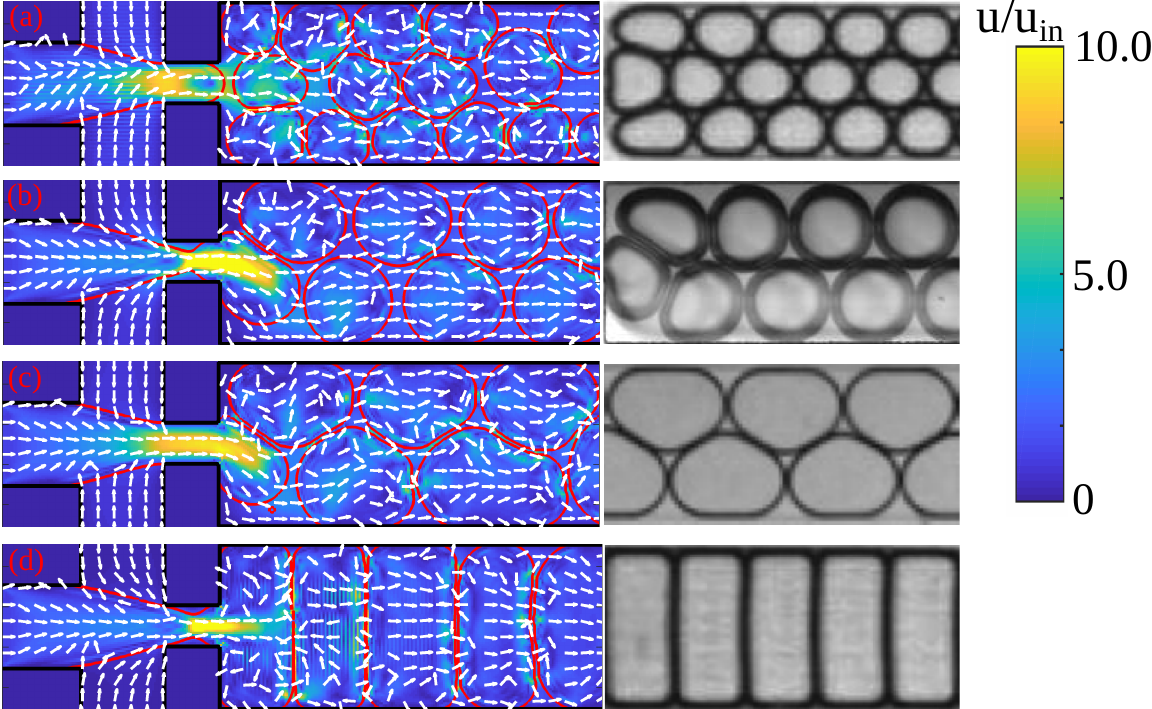}
\caption{\label{fig2} Emulsion structures in microchannels. Different patterns are obtained by tuning the dispersed-to-continuous inlet flow ratio $\phi=u_d/2u_c$, (a) Hex-Three, $\phi=1$, (b) Wet Hex-Two, $\phi=1.5$, (c) Dry Hex-Two, $\phi=2$, (d) Hex-One, $\phi=3.6$. Simulations (left column) are compared with experiments (right column) of Marmottant et al. \cite{raven2009microfluidic}. Continuous red lines indicate droplet interfaces, while white arrows denote the direction of the velocity field. The color map defines the magnitude of the velocity field, from blue (low) to yellow (high), where $u_{in}$ is the speed measured at the inlet channel.}
\end{figure}

In Ref.\cite{marmottant2009microfluidics}, Marmottant {\it et al.} show that, by merely turning up the pressure of the dispersed phase, a variety of ordered hexagonal emulsion patterns can be obtained and assembled in the outlet channel.  The typical structures are shown in Fig.\ref{fig2} (experiments on the right panel), and are named according to the number of droplet rows in the channel: going from the top to the bottom we have hex-three, hex-two and hex-one (also called ``bamboo''). 

In the left panel of Fig.\ref{fig2}, we report the different flowing crystalline-like structures observed in our simulations upon varying $\phi$ and keeping ${\cal N}_c=0.5$. While, for $\phi=1$ (Fig.\ref{fig2}a), a regular array of mono-disperse droplets arranged in a three-row structure (Hex-three phase) is produced,  increasing $\phi$ significantly affects this picture. When $\phi=1.5$ (Fig.\ref{fig2}b), the resulting emulsion phase consists, once more, of monodisperse droplets, but now uniformly accommodated along two parallel rows (Wet Hex-Two phase), and an analogous arrangement is observed when $\phi=2$ (Fig.\ref{fig2}c), although here larger droplets exhibit a wider interface deformation. Finally, when $\phi=3.6$ (Fig.\ref{fig2}d), the system displays a single-row structure, in which highly packed brick-like shaped droplets arrange in a single-file pattern (Hex-One phase).        

The visual match between simulations and experiments is clearly impressive.

\begin{figure}
\includegraphics[scale=0.47]{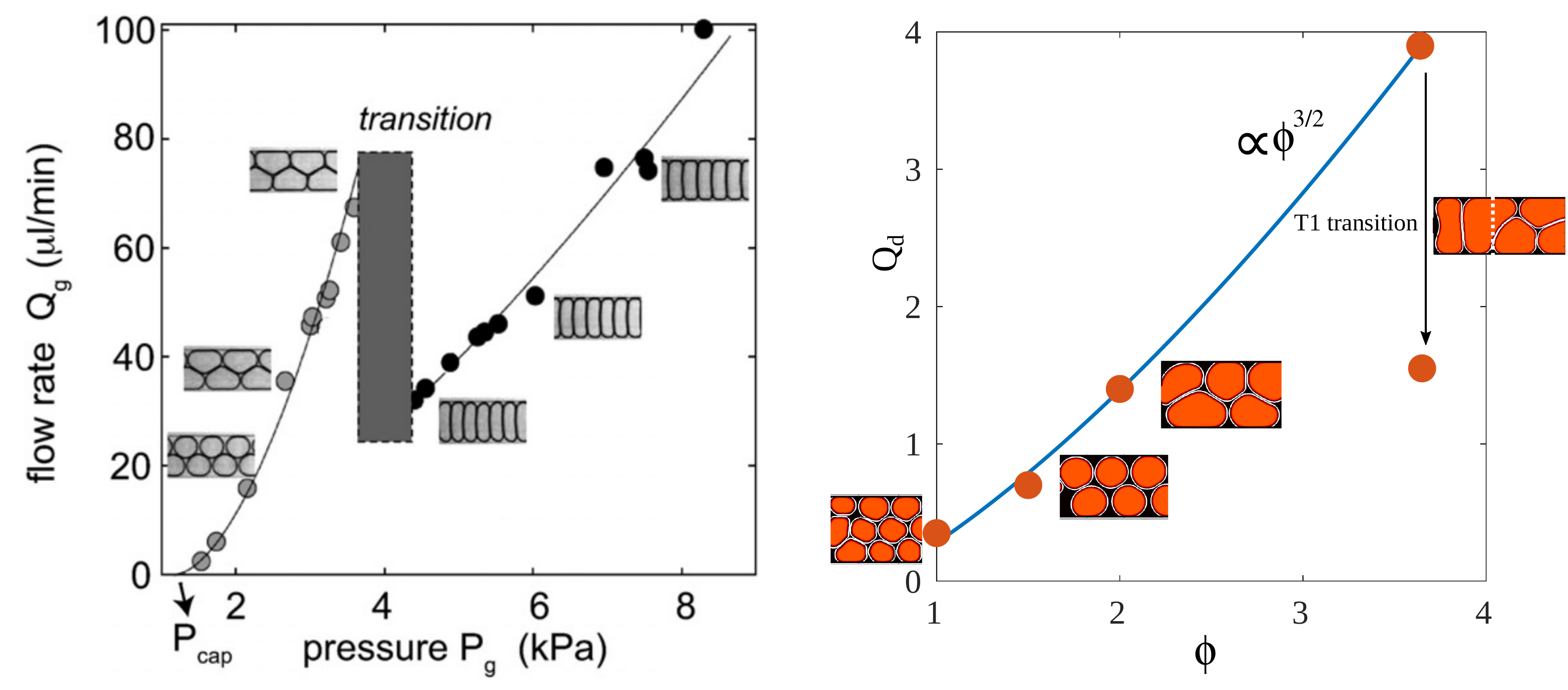}
\caption{\label{fig3} Left: The phase diagram of a flowing microfluidic foam of Ref.\cite{marmottant2009microfluidics}. $P_g$ and $Q_g$ are the gas pressure and gas flow rate, respectively. For low values of $P_g$, the flow rate scales as $Q_g\propto P^{1/\alpha}$, where $\alpha\simeq 2/3$. At $P_g\simeq 4$, a pronounced decrease of $Q_g$ marks a sharp transition from the hex-two phase to the hex-one.  Right:  Phase diagram of a soft flowing crystal, where $\phi=u_d/2u_c$ and $Q_d=\nu V_d$ is the flow rate of the disperse phase, computed as the product of the frequency $\nu$ of droplet generation at the nozzle times the volume $V_d $ of the generated droplet. For low values of $\phi$ the flow rate scales as power law $Q_d\propto \phi^{\beta}$, where $\beta\simeq 3/2$. The non-equilibrium phase transition from the hex-three phase to the hex-two phase occurs when $\phi\simeq 1.5$. When $\phi\simeq 3.5$, a dramatic decrease of $Q_d$ marks the transition to the hex-one phase. The white dotted line indicates the transition front.}
\end{figure}

The formation and the arrangement of these regular patterns are decisively affected by the flow rate. Indeed, it has been shown that, for a gas-liquid foam, the flow rate is a power-law function of the applied pressure of the gas set at the inlet channel, i.e. $Q_f\propto P_g^{1/\alpha}$ where $\alpha\simeq 2/3$ \cite{cantat2004,marmottant2009microfluidics}, as long as $P_g$ is sufficiently small (see Fig.~\ref{fig3}, left). In particular, increasing $P_g$, one moves from an hex-two phase, made of two rows of rather spherical droplets, to a closer packed state in which larger bubbles occupy the whole microchannel. Afterwards, a sharp decrease of $Q_g$, due to the augmented friction of droplet interfaces with the solid walls, indicates a non-equilibrium order-order transition towards the hex-one phase.

Remarkably, such dynamic behaviour is fully captured by our simulations. In Fig.~\ref{fig3} (right), we show a phase diagram of $Q_d=\nu V_d$ as a function of $\phi$, where $\nu$ is the frequency of droplet generation and $V_d$ is the volume of the generated droplet. Here the flow ratio $\phi$ plays the same role of the gas pressure $P_g$ and is fixed at the inlet channel. In excellent agreement with the flow curve shown in Ref.~\cite{marmottant2009microfluidics}, the flow rate $Q_d$ scales as a power law of $\phi^{1/\beta}$, where $\beta$ is approximately $2/3$. Now, by increasing $\phi$ one switches from the hex-three phase to the hex-two phase, the last stable as long as $\phi\simeq 3.5$. Like the previous case, further augmenting $\phi$ produces a dramatic reduction of $Q_d$, which signs that transition to the hex-one phase. Such transition is triggered by a topological rearrangement, known as T1 event \cite{weaire1999, raven2009microfluidic} (see the white dotted line in Fig.\ref{fig3}, left), which corresponds to an exchange in neighbouring droplets to diminish the energy of the fluid interface. When this event occurs, two different phases, hex-two and hex-one, simultaneously coexist, separated by a transition front that breaks the homogeneity of the pattern.

Clearly, the flow field plays a fundamental role in a soft flowing crystal. It is therefore of particular relevance to investigate more carefully its structure in order to assess how it affects shape and stability of the flowing crystal and, potentially, its final design.

In the following section, we discuss the flow structure characterising different flowing crystals in detail.

\subsection{Fluid-structure interaction and stability of the crystal pattern}\label{R3}

\begin{figure}
\includegraphics[scale=0.7]{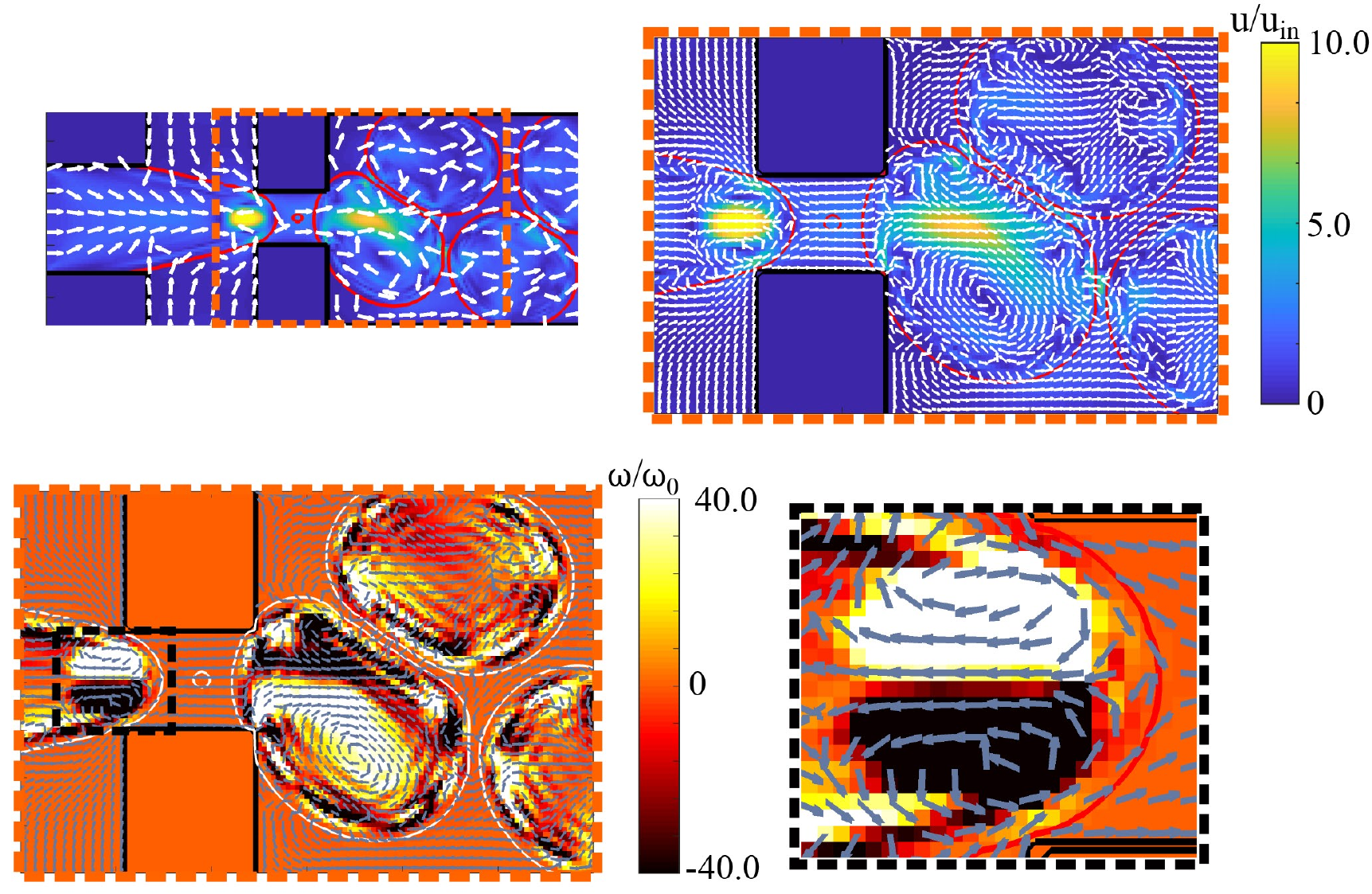}
\caption{\label{fig4} Right: Instantaneous velocity field (top row) and vorticity fields (bottom row) after the break-up stage. Left: Zoom of the fields
within the orifice and in its surroundings. Strong recirculations form both within the droplets and at the tip of the retracting jet. $\omega_0$ is a
characteristic pulsation computed as $\omega_0=u_0/L$, where $u_0$ is the speed of the dispersed phase at the inlet channel and $L$ is the length of the
channel.}
\end{figure}

In all cases described so far, the velocity field looks rather uniform upstream, in the inlet channels, while it is significantly affected by the droplet arrangement downstream, within the outlet channel (see Fig.~\ref{fig2}). 
Indeed, the chaotic-like pattern observed at $\phi=1$, progressively turns into a more uniform and almost unidirectional one as $\phi$ augments, due to the increasing droplet size and the concurrent loss of interfacial area. 
In particular, at high values of $\phi$, the non-uniform velocity field remains mainly confined nearby the droplet 
interfaces, whereas, within each droplet, it points rightwards almost everywhere.

At a closer inspection, however,  the coupling between hydrodynamics and interface dynamics reveals a number of highly non-trivial aspects. 

In Fig.~\ref{fig4} we show a zoom of the velocity field (top) and of the vorticity $\omega$ (bottom) 
when $\phi=1.5$. Interestingly, at the entrance of the nozzle and within the inlet jet, two strong counter-rotating 
vortices appear. Together with the fluid flow of the continuous phase, they promote the squeezing of the interface and its further elongation within the orifice, before the droplet break-up. 
Afterwards, the flow exhibits an {\it apparently} chaotic pattern near and within the fluid interfaces, whereas 
it exhibits a more regular and approximately uniform pattern in the bulk of the droplets and outside, far from the interfaces. 
Intense fluid recirculations, rotating either clockwise or anticlockwise, still occur within the generated 
droplets near the nozzle (where the fluid flow magnitude is larger than the rest of the system), and progressively disappear when the droplets move forward within the outlet channel. 

The above pictures prompts a number of questions, such as, how robust is this scenario? 
Is the feasibility of the material guaranteed by solely monitoring the physics at the device scale or may effects occurring at the interface level potentially destabilize the material? 

In Ref.\cite{tiribocchipre2019} we have described how near-contact forces, typically showcasing in a film of fluid formed by the interfaces, could affect the mechanical properties of the SFC. 
Their effect is modelled via a mesoscopic repulsive force $\vec{F}_{rep}$, which competes with surface tension (capillary forces) to inhibit droplet coalescence and coarsening of the material. 
We have shown that, if the near-contact forces are comparable or exceed the effects of surface tension (i.e. when ${\cal N}_c\sim 1$, or larger), counter-rotating micro-vorticity patterns emerge
within the film of fluid. These promote the formation of ripples propagating along with the interface that may eventually destabilize the film and potentially lead to the rupture of the material.

\section{Why it works? Extended Universality}

At the end of this work, a question stands tall: how could the mesoscale approach used throughout this paper possibly work and in a close-to-quantitative form, in the face of very substantial coarse-graining, basically two orders of magnitude in space?

We maintain that the answer traces to a sort of ``Extended Universality'' (EU) of the underlying physics in point, a {\it sine-qua non} for the mesoscale approach to be effective at all.
If universality would hold,  a continuum approach would do, if it were lost
altogether, molecular dynamics would be the only option.

Universality means dependence on dimensionless groups rather than the specific values
of the competing forces and interactions. 
For the case of near contact interactions, besides the capillary number, such group can be identified with the ratio of near-contact versus capillarity forces.
We have not found any existing name for such number and we refrain from introducing
a new one ourselves, so we simply call it the ``near-contact-number''  ${\cal N}_c$.

Another relevant dimensionless group is the Cahn number, defined as the ratio of the
film thickness to the droplet diameter, $Cn = h/D$. 

As long as the simulation can feature the same ${\cal N}_c$ and $Cn$ as the physical problem, success
can be expected in predicting the main features of the phenomenon.
This is standard universality. 

In our simulations ${\cal N}_c \sim 0.1$ which is indeed comparable 
with the physical value \cite{israelnature,israelpnas}.
However, the simulated Cahn number is $Cn \sim 0.1$, three orders
of magnitude above the physical value. 

How come it still works?

At this point, it is important to realise that the mesoscale approach can also work under less restrictive conditions, 
namely that the simulated value of a dimensional  
number be different from the physical one, in our case:
\begin{equation}\label{UNI}
Cn_{sim} \ne  Cn_{phys}
\end{equation}

Provided the error due to such mismatch remains negligible as compared
to the physical value of the observable under inspection, the basic physics
remains unscathed. This is what we mean by ``Extended Universality'' (EU).

On general grounds, EU applies whenever i) the physical value of the dimensionless group is well
below 1, and ii) the mismatch error exhibits a smooth dependence on the number in point, say $\lambda$,
in equations
\begin{equation}
\label{SBU}
\frac{\delta O}{O} = c_1  \lambda + c_2 \lambda^2 + \dots,
\end{equation}
where $\delta O$ is the mismatch error on the physical observable $O$ 
and $c_1$ and $c_2$ are $O(1)$ numerical constants. 
 
A typical example is the Mach number in LB simulations of slow flows, say porous media.
The physical Mach number is of the order of $M \sim 10^{-4}$ or less, which would be totally unpractical in LB simulations, due to unmanageably small time steps.
A standard trick is to raise the Mach number artificially, say 
$M \sim 0.1$, gaining three orders of magnitude in the process.
This comes at the cost of a compressibility error, but since compressibility 
effects scale like the Mach number squared, the simulated fluid
carries density fluctuations of the order of $10^{-2}$, instead 
of $10^{-8}$ in the physical flow. This is a huge relative error, but still
tolerable as an absolute one: for many problems 1 per cent error does not obscure the basic 
physics, hence it can  be tolerated without major disruptions, providing dramatic computational savings 
in the process.

We may call expression (\ref{UNI}) a {\it soft-constraint}, namely one which tolerates significant and sometimes even dramatic departures from strict universality, $Cn_{sim}=Cn_{phys}$, without causing appreciable blurring of the physical picture. Or, differently restated, the physics is largely insensitive to the Cahn number; for droplets 100 microns in size, it makes little difference whether their interstitial film
is 1 micron instead of 10 nanometers thick. Once the ratio $h/D$ is well below 1, it does not matter much whether it is just well below one or many orders of magnitude bellow.
We maintain that, just like the Mach number in porous media, the Cahn number falls precisely in 
the forgiving class of Extended Universality.

Extended Universality is a precious gift which permits to save orders of magnitude
in computational demand. A gift that can never be given for granted {\it a priori}, but rather needs to be checked case-by-case. 
Incidentally, it bears a definite relation to the notion of ``sloppiness'', namely 
the property of many physical and especially biological systems, to maintain their behaviour largely unaffected even under major changes of the so-called ``sloppy'' parameters
\cite{gutenkunst}, i.e. parameters which are largely unconstrained towards experimental ones.
The crucial difference, though, is that sloppy parameters are usually many and exposed
to subjective modelling choices, whereas Extended Universality refers to a few dimensionless numbers
strictly dictated by the competing physical mechanism under inspection.    

Perhaps, the most far-reaching merit of the mesoscopic LB approach discussed in this paper
is precisely to provide an efficient ``tool of discovery'', for spotting Extended Universality
wherever it happens to hold.   

\section{Conclusions and outlook}

The mesoscale method discussed in this paper delivers several non-trivial insights into the physics of near-contact interfaces of direct relevance to the rheology of soft crystals.

First, it shows that, although overly complex and non-linear, such rheology is amenable to substantial coarse-graining, basically two orders of magnitude, from 
10 nm to 1 micron, thanks to the insensitivity to the Cahn number, a property we have dubbed Extended Universality. This allows a dramatic savings of computational resources. 

Second, it highlights a remarkable degree of extended universality in the 
physics of near-contact interfaces: once a mechanism for preventing 
coalescence is appropriately secured, and its strength
properly calibrated, nanoscale details do not appear to be essential.
Yet, not any mesoscopic treatment would do: for instance, lattice potentials such as 
those currently in use in the LB literature \cite{benzi2009mesoscopic,sbragaglia2012emergence,fei2018mesoscopic}, are too short-ranged 
to capture the complex physics discussed in this paper.

Third, the simulations highlight a very fascinating interplay between 
micro-hydrodynamic recirculation patterns and the interface dynamics.
Near-contact interactions affect the interface fluid, and, once sufficiently intense,
they trigger micro-vorticity patterns, which in turn affect the NCIs through the shape changes of the interface. 
This non-linear and self-consistent fluid-structure coupling remains to be 
explored in-depth, but many interesting questions for future research 
can be safely be anticipated.

Among others: How fast can the soft crystal flow before it melts or ruptures? 
Does the self-consistent coupling between hydrodynamic vorticity and
interface dynamics permit to fine-tune the design of new porous mesoscale
materials with wiggly interfaces?
Under what parameter regimes, if any, do soft-flowing crystals connect to Time Crystals,
i.e. namely coherent patterns supporting periodicity in both space and time \cite{wilczekprl,goldstein_pt}?

We hope and expect that the mesoscale LB method presented in this work may help gaining new and fresh insights into the fascinating questions above.

\section*{Materials and Methods}

In this section we shortly summarize the numerical model employed in our simulations.
It is a variant of the Lattice Boltzmann (LB) method for multiphase flows (for a comprehensive review on LB see \cite{succi2018lattice,kruger2017lattice,sukop2005lattice,benzi1992}), based on the color-gradient model of Leclaire et al.\cite{leclaire2012,leclaire2017}, and augmented with a repulsive forcing term accounting for the effects of near-contact interactions occurring at the fluid interface level \cite{montessori2019jfm}. Such method has been successfully adopted to simulate, for instance, the physics of dense fluid emulsions in microchannels \cite{montessori2019mesoscale,montessori2019jfm}.

In this model the two components of the binary fluid are described by two distinct sets of distribution functions $f_i^k$ ($k=1,2$), whose evolution is governed by a discrete Boltzmann equation of the form
\begin{equation} \label{CGLBE}
f_{i}^{k} \left(\vec{x}+\vec{c}_{i}\Delta t,\,t+\Delta t\right) =f_{i}^{k}\left(\vec{x},\,t\right)+\Omega_{i}^{k}[ f_{i}^{k}\left(\vec{x},\,t\right)].
\end{equation}
$f_i^k$ represents the probability of finding a particle of the $k$th component at position $\vec{x}$ and time $t$ with discrete velocity $\vec{c}_{i}$, and $i$ is an index running over the lattice discrete directions $i=0,...,b$, where $b=26$ for the three dimensional 27 speed lattice vectors (D3Q27) \cite{succi2018lattice}. The time step $\Delta t$ is kept fixed to $1$, as common in LB simulations \cite{succi2018lattice}.

The local density $\rho^{k}(\vec{x},t)$ of the $k$th component and the total momentum of the mixture $\rho\vec{u}=\sum_k\rho^{k}\vec{u^k}$ are given by the zeroth and the first order moment of the  distribution functions, i.e.  $\rho^{k}(\vec{x},\,t) = \sum_i f_{i}^{k}(\vec{x},\,t)$ and $\rho \vec{u} = \sum_i  \sum_k f_{i}^{k}\left(\vec{x},\,t\right) \vec{c}_{i}$.

The last term of Eq.(\ref{CGLBE}) is the collision operator, which consists of three terms \cite{gunstensen1991,leclaire2012,leclaire2017}
\begin{equation}
\Omega_{i}^{k} = \left(\Omega_{i}^{k}\right)^{(3)}\left[\left(\Omega_{i}^{k}\right)^{(1)}+\left(\Omega_{i}^{k}\right)^{(2)}\right].
\end{equation}

In the above, $\left(\Omega_{i}^{k}\right)^{(1)}$ represents the standard relaxation contribution \cite{succi2018lattice}, which sets the kinematic viscosity $\nu$ of the mixture. The second term $\left(\Omega_{i}^{k}\right)^{(2)}$ is the perturbation step \cite{gunstensen1991}, accounting for the interfacial tension.  Finally, $\left(\Omega_{i}^{k}\right)^{(3)}$ is the recoloring step \cite{gunstensen1991,latva2005}, which favours the segregation between the two species so as to minimise their mutual diffusion.

By performing a Chapman-Enskog expansion of the distribution functions \cite{benzi1992,chen1998,carenza2019}, it can be shown that the hydrodynamic limit of Eq.(\ref{CGLBE}) converges to the continuity and the Navier-Stokes equations
\begin{equation}\label{cont}
 \frac{\partial\rho}{\partial t} + \nabla\cdot \rho\vec{u}=0
\end{equation}
\begin{equation}\label{Nav_stok}
  \frac{\partial \rho \vec{u}}{\partial t}+\nabla\cdot\rho\vec{u}\vec{u}=-\nabla p+\nabla\cdot[\rho\nu(\nabla\vec{u}+\nabla\vec{u}^T)]+\nabla\cdot\mathbf{\Sigma},
\end{equation}
where $p$ is the ideal gas pressure and $\mathbf{\Sigma}$ is the capillary stress tensor, given by
\begin{equation}
\mathbf{\Sigma}= \frac{\sigma}{2 |\nabla \rho|}(|\nabla \rho|^2\mathbf{I} - \nabla \rho \times \nabla \rho).
\end{equation}
Here $\sigma$ is the surface tension \cite{succi2018lattice,kruger2017lattice}.

\begin{figure}
\includegraphics[scale=0.65]{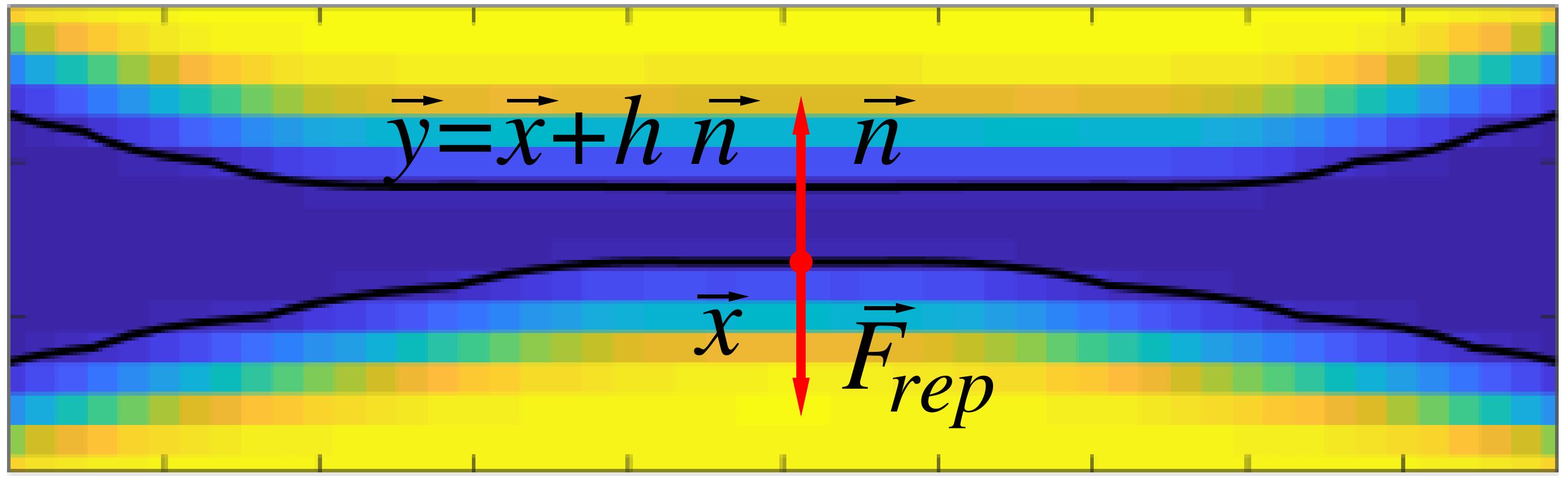}
\caption{\label{fig5} Mesoscale representation of near contact forces between two immiscible fluid droplets.
 $\vec{F}_{rep}$ is the repulsive force and $\vec{n}$ is the unit vector perpendicular to the fluid interface. $\vec{x}$ and $\vec{y}$ indicate
the positions, placed a distance $\vec{h}$, located within the fluid interface.}\label{sketchrep}
\end{figure}

The stress jump across a fluid interface is related to $\sigma$ by the following expression
\begin{equation}\label{stjump}
 \mathbf{T}^1\cdot \vec{n} - \mathbf{T}^2 \cdot \vec{n}=-\nabla(\sigma \mathbf{I} - \sigma (\vec{n}\otimes \vec{n})) - \pi \vec{n},
\end{equation}
where $\mathbf{T}^k=-p_k\mathbf{I}+\rho_k\nu(\nabla\vec{u}+\nabla\vec{u}^T)$ is the stress tensor of the $k$th component, 
$\vec{n}=-\nabla\psi/|\nabla\psi|$ (with  $\psi=(\rho^1-\rho^2)/(\rho^1+\rho^2)$ scalar order parameter ranging between 
$-1$ and $1$) is a unit vector normal to the interface, and $\pi[h(\vec{x})]$ represents the repulsion between 
neighboring fluid interfaces, aimed at providing a mesoscale representation of all the repulsive near-contact 
forces (i.e., Van der Waals, electrostatic, steric and hydration repulsion) acting on much smaller scales ($\sim  O(1 \; nm)$) 
than those resolved on the lattice (typically well above hundreds of nanometers). 
The term $h(\vec{x})$ sets the distance between sites $\vec{x}$ and $\vec{y}=\vec{x}+ h(\vec{x}) \vec{n}$ along 
the normal $\vec{n}$ at the two interfaces (see Fig. \ref{sketchrep}). 

By neglecting variations of the surface tension along the interface, one can approximate  $\mathbf{T}\simeq -p\mathbf{I}$ (\cite{brackbill1992}) and, by further projecting Eq.(\ref{stjump}) along the normal to the interface, the augmented Young-Laplace equation is obtained (\cite{chan2011film,williams1982})
\begin{equation}
p_2 - p_1=\sigma (\nabla \cdot \vec{n}) - \pi
\end{equation}

The additional term $\pi$ can be included within the LB framework by adding a 
forcing term acting only at the fluid interfaces in near contact. This is given by
\begin{equation}\label{near_c}
\vec{F}_{rep}= \nabla \pi := - A_{h}[h(\vec{x})]\vec{n} \delta_I,
\end{equation}

where $\delta_I=\frac{1}{2}|\nabla\psi|$ is a function (computed by using a second order discretization of the gradint 
operator~\cite{thampi2013isotropic}) confining the near-contact effects at the fluid interface, and $A_h[h(\vec{x})]$ 
controls the strength (force per unit volume) of the near-contact interactions. 
It is set equal to a constant $A$ if $h\leq h_{min}$, it decays as $h^{-3}$ if $h_{min}<h\leq h_{max}$ and 
it is equal to zero if $h>h_{max}$. 
In our simulations $h_{min}=2$ and $h_{max}=4$ lattice spacings. Although other functional forms are plausible, 
this one is overall sufficient to avoid droplet coalescence and, more importantly, to correctly describe the 
physics at different lengthscales. From a computational standpoint, the (weak) non-local nature 
of the forcing term may affect the parallel efficiency of the method when implemented on distributed memory systems. 
Such issue, absent in a distributed memory architecture (used in this simulations), could be partially circumvented 
by a domain decomposition with an appropriate number of halo nodes.

The repulsive force is finally added in Eq.\ref{CGLBE} (by using the exact difference method \cite{kupershtokh2010criterion}) 
and is applied only to the dispersed phase. This modifies the stress tensor of the Navier-Stokes equation which can be 
recovered through the substitution $\nabla\cdot\mathbf{\Sigma}\rightarrow\nabla\cdot(\mathbf{\Sigma}+\pi\mathbf{I})$.

\vskip 1cm

\section*{Acknowledgements}
The authors acknowledge funding from the European Research Council under the European Union's Horizon 2020 Framework
Programme (No. FP/2014-2020) ERC Grant Agreement No.739964 (COPMAT).

%\bibliographystyle{elsarticle-num}
%\bibliography{bibliography}

\begin{thebibliography}{10}
\expandafter\ifx\csname url\endcsname\relax
  \def\url#1{\texttt{#1}}\fi
\expandafter\ifx\csname urlprefix\endcsname\relax\def\urlprefix{URL }\fi
\expandafter\ifx\csname href\endcsname\relax
  \def\href#1#2{#2} \def\path#1{#1}\fi

\bibitem{fernandez2016}
A.~Fernandez-Nieves, A.~M. Puertas, Fluids, Colloids and Soft Materials: An
  Introduction to Soft Matter Physics, Wiley, New York, US, 2016.

\bibitem{piazza2012}
R.~Piazza, Soft Matter: The stuff that dreams are made of, Springer, Rotterdam,
  Netherlands, 2012.

\bibitem{lavren2003}
M.~Kleman, O.~D. Lavrentovich, J.~Friedel, Soft Matter Physics: An
  Introduction, Springer Verlag, 2003.

\bibitem{jones2002}
R.~A.~L. Jones, Soft Condensed Matter, Oxford University Press, Oxford, 2002.

\bibitem{degennes1994}
P.~J. De~Gennes, J.~Prost, The Physics of Liquid Crystals, Oxford University
  Press, Oxford, 1994.

\bibitem{bernaschi2019}
M.~Bernaschi, S.~Melchionna, S.~Succi, Mesoscopic simulations at the
  physics-chemistry-biology interface, Rev. Mod. Phys. 91 (2019) 025004.

\bibitem{schiller2018}
U.~D. Schiller, T.~Kr\"uger, O.~Henrich, Mesoscopic modelling and simulation of
  soft matter, Soft Matter 14 (2019) 9--26.

\bibitem{henrich2018}
O.~Henrich, Y.~A. Gutierrez~Fosado, T.~Curk, T.~E. Ouldridge, Coarse-grained
  simulation of dna using lammps, Eur. Phys. J. E 41 (2018) 57.

\bibitem{prosperetti2007}
A.~Prosperetti, G.~Trygvasson, Computational Methods for Multiphase Flows,
  Cambridge University Press, UK, 2007.

\bibitem{kartunnen2009}
M.~Kartunnen, I.~Vattulainen, A.~Lukkarinen, Novel Methods in Soft Matter
  Simulations, Springer, New York, US, 2009.

\bibitem{kruger2013numerical}
T.~Kr{\"u}ger, S.~Frijters, F.~G{\"u}nther, B.~Kaoui, J.~Harting, Numerical
  simulations of complex fluid-fluid interface dynamics, The European Physical
  Journal Special Topics 222~(1) (2013) 177--198.

\bibitem{fili98}
O.~Filippova, D.~H\"anel, Grid refinement for lattice-bgk models, J. Comput.
  Phys. 147 (1998) 219--228.

\bibitem{stein2017}
M.~O. Steinhauser, Multiscale Modeling of Fluids and Solids - Theory and
  Applications, Springer, 2017.

\bibitem{dupuis2007}
A.~Dupuis, E.~M. Kotsalis, P.~Koumoutsakos, Coupling lattice boltzmann and
  molecular dynamics models for dense fluids, Phys. Rev. E 75 (2007) 046704.

\bibitem{distaso2016}
G.~Di~Staso, H.~J.~H. Clercx, S.~Succi, F.~Toschi, Dsmc-lbm mapping scheme for
  rarefied and non-rarefied gas flows, J. Comp. Sci. 17 (2016) 357--369.

\bibitem{bernaschi2009}
M.~Bernaschi, S.~Melchionna, S.~Succi, M.~Fyta, E.~Kaxiras, J.~K. Sircar,
  Muphy: A parallel multi physics/scale code for high performance bio-fluidic
  simulations, Comp. Phys. Comm. 180 (2009) 1495--1502.

\bibitem{succi2018lattice}
S.~Succi, The Lattice Boltzmann Equation: For Complex States of Flowing Matter,
  Oxford University Press, 2018.

\bibitem{yeomans2006}
J.~M. Yeomans, Mesoscale simulations: Lattice boltzmann and particle
  algorithms, Physica A: Statistical Mechanics and its Applications 369 (2006)
  159--184.

\bibitem{montessori2019jfm}
A.~Montessori, M.~Lauricella, N.~Tirelli, S.~Succi, Mesoscale modeling of
  near-contact interactions for complex flowing interfaces, Jour. Fluid Mech.
  872 (2019) 327--347.

\bibitem{montessori2019mesoscale}
A.~Montessori, M.~Lauricella, S.~Succi, Mesoscale modelling of soft flowing
  crystals, Phil. Trans. Roy. Soc., Ser. A 377 (2019) 20180149.

\bibitem{raven2009microfluidic}
J.-P. Raven, P.~Marmottant, Microfluidic crystals: dynamic interplay between
  rearrangement waves and flow, Physical review letters 102~(8) (2009) 084501.

\bibitem{marmottant2009microfluidics}
P.~Marmottant, J.-P. Raven, Microfluidics with foams, Soft Matter 5~(18) (2009)
  3385--3388.

\bibitem{vecchiolla2019dislocation}
D.~Vecchiolla, S.~L. Biswal, Dislocation mechanisms in the plastic deformation
  of monodisperse wet foams within an expansion--contraction microfluidic
  geometry, Soft matter 15~(30) (2019) 6207--6223.

\bibitem{fei2020discrete}
L.~Fei, A.~Scagliarini, K.~H. Luo, S.~Succi, Discrete fluidization of dense
  monodisperse emulsions in neutral wetting microchannels, Soft Matter (2020).

\bibitem{duffy1998}
D.~C. Duffy, J.~C. McDonald, O.~J.~A. Schueller, J.~M. Whitesides, Rapid
  prototyping of microfluidic systems in poly(dimethysiloxane), Anal. Chem. 70
  (1998) 4974--4984.

\bibitem{anna2007}
J.~F. Christopher, S.~L. Anna, Microfluidic methods for generating
  continuousdroplet streams, J. Phys. D. Appl. Phys. 40 (2007) R319.

\bibitem{garstecki2004}
P.~Garstecki, I.~Gitlin, W.~DiLuzio, G.~M. Whitesides, E.~Kumacheva, H.~A.
  Stone, Formation of monodisperse bubbles in a microfluidic flow-focusing
  device, Appl. Phys. Lett. 85 (2004) 2649--2651.

\bibitem{garstecki2006flowing}
P.~Garstecki, G.~M. Whitesides, Flowing crystals: nonequilibrium structure of
  foam, Phys. Rev. Lett. 97 (2006) 024503.

\bibitem{squires2005}
T.~M. Squires, S.~R. Quake, Microfluidics: Fluid physics at the nanoliter
  scale, Rev. Mod. Phys. 77 (2005) 977.

\bibitem{dolletprl}
B.~Dollet, W.~van Hoeve, J.~P. Raven, P.~Marmottant, M.~Versluis, Role of the
  channel geometry on the bubble pinch-off in flow-focusing devices, Phys. Rev.
  Lett. 100 (2008) 034504.

\bibitem{garstecki2005}
P.~Garstecki, H.~A. Stone, G.~M. Whitesides, Mechanism for flow-rate controlled
  breakup in confined geometries: A route to monodisperse emulsions, Phys. Rev.
  Lett. 94 (2005) 164501.

\bibitem{montessoriprf}
A.~Montessori, M.~Lauricella, S.~Succi, E.~Stolovicki, D.~Weitz, Elucidating
  the mechanism of step emulsification, Phys. Rev. F. 3 (2018) 072202.

\bibitem{garsteckiprl2005}
P.~Garstecki, M.~J. Fuerstman, G.~M. Whitesides, Nonlinear dynamics of a
  flow-focusing bubble generator: An inverted dripping faucet, Phys. Rev. Lett.
  94 (2005) 234502.

\bibitem{kimmins2010}
S.~D. Kimmins, N.~R. Cameron, Functional porous polymers by emulsion
  templating: Recent advances, Adv. Funct. Mater. 21 (2011) 211--225.

\bibitem{costantini2014}
M.~Costantini, C.~Colosi, J.~Guzowski, A.~Barbetta, J.~Jaroszewicz,
  W.~Swieszkowski, M.~Dentini, P.~Garstecki, Highly ordered and tunable
  polyhipes by using microfluidics, J. Mater. Chem. B 2 (2014) 2290--2300.

\bibitem{robinson2014}
J.~L. Robinson, R.~S. Moglia, M.~C. Stuebben, M.~A.~P. McEnery,
  E.~Cosgriff-Hernandez, Achieving interconnected pore architecture in
  injectable polyhipes for bone tissue engineering, Tissue Eng. Part. A 20
  (2014).

\bibitem{tiribocchipre2019}
A.~Tiribocchi, A.~Montessori, S.~Miliani, M.~Lauricella, M.~La~Rocca, S.~Succi,
  Microvorticity fluctuations affect the structure of thin fluid films, Phys.
  Rev. E 100 (2019) 042606.

\bibitem{derjaguin1940}
B.~V. Derjaguin, On the repulsive forces between charged colloid particles and
  on the theory of slow coagulation and stability of lyophobe sols, Trans.
  Faraday Soc. 36 (1940) 730.

\bibitem{cantat2004}
I.~Cantat, N.~Kern, R.~Delannay, Dissipation in foam flowing through narrow
  channels, Europhys. Lett. 65 (2004) 726--732.

\bibitem{weaire1999}
D.~Weaire, S.~Hutzler, The physics of foams, Oxford University Press, Oxford,
  1999.

\bibitem{israelnature}
J.~N. Israelachvili, R.~Pashley, The hydrophobic interaction is long range,
  decaying exponentially with distance, Nature 300 (1982) 341--342.

\bibitem{israelpnas}
S.~H. Donaldson, C.~Ted~Lee, B.~F. Chmelka, J.~N. Israelachvili, General
  hydrophobic interaction potential for surfactant/lipid bilayers from direct
  force measurements between light-modulated bilayers, Proc. Nat. Acad. Sci.,
  USA 108 (2011) 15699--15704.

\bibitem{gutenkunst}
R.~N. Gutenkunst, J.~J. Waterfall, F.~P. Casey, K.~S. Brown, C.~R. Myers, J.~P.
  Sethna, Universally sloppy parameter sensitivities in systems biology models,
  Plos Comput. Biol. 3 (2007) e189.

\bibitem{benzi2009mesoscopic}
R.~Benzi, S.~Chibbaro, S.~Succi, Mesoscopic lattice boltzmann modeling of
  flowing soft systems, Physical review letters 102~(2) (2009) 026002.

\bibitem{sbragaglia2012emergence}
M.~Sbragaglia, R.~Benzi, M.~Bernaschi, S.~Succi, The emergence of
  supramolecular forces from lattice kinetic models of non-ideal fluids:
  applications to the rheology of soft glassy materials, Soft Matter 8~(41)
  (2012) 10773--10782.

\bibitem{fei2018mesoscopic}
L.~Fei, A.~Scagliarini, A.~Montessori, M.~Lauricella, S.~Succi, K.~H. Luo,
  Mesoscopic model for soft flowing systems with tunable viscosity ratio,
  Physical Review Fluids 3~(10) (2018) 104304.

\bibitem{wilczekprl}
F.~Wilczek, Quantum time crystals, Phys. Rev. Lett. 109 (2012) 160401.

\bibitem{goldstein_pt}
R.~Goldstein, Coffee stains, cell receptors, and time crystals: Lessons from
  the old literature, Physics Today 71 (2018) 32.

\bibitem{kruger2017lattice}
T.~Kr{\"u}ger, H.~Kusumaatmaja, A.~Kuzmin, O.~Shardt, G.~Silva, E.~M. Viggen,
  The lattice Boltzmann method, Springer, 2017.

\bibitem{sukop2005lattice}
M.~C. Sukop, D.~T. Thorne, Lattice Boltzmann modeling: An Introduction for
  Geoscientists and Engineers, Springer, 2005.

\bibitem{benzi1992}
R.~Benzi, S.~Succi, M.~Vergassola, The lattice boltzmann equation: theory and
  applications, Phys. Rep. 222~(3) (1992) 145--197.

\bibitem{leclaire2012}
S.~Leclaire, M.~Reggio, J.-Y. Tr{\'e}panier, Numerical evaluation of two
  recoloring operators for an immiscible two-phase flow lattice boltzmann
  model, Appl. Math. Model. 36 (2012) 2237--2252.

\bibitem{leclaire2017}
S.~Leclaire, A.~Parmigiani, O.~Malaspinas, B.~Chopard, J.~Latt, Generalized
  three-dimensional lattice boltzmann color-gradient method for immiscible
  two-phase pore-scale imbibition and drainage in porous media, Phys. Rev. E 95
  (2017) 033306.

\bibitem{gunstensen1991}
A.~K. Gunstensen, D.~H. Rothman, S.~Zaleski, G.~Zanetti, Lattice boltzmann
  model of immiscible fluids, Phys. Rev. A 43 (1991) 4320.

\bibitem{latva2005}
M.~Latva-Kokko, D.~H. Rothman, Diffusion properties of gradient-based lattice
  boltzmann models of immiscible fluids, Phys. Rev. E 71 (2005) 056702.

\bibitem{chen1998}
S.~Chen, G.~D. Doolen, Lattice boltzmann method for fluid flows, Annu. Rev.
  Fluid. Mech. 30 (1998) 329--364.

\bibitem{carenza2019}
L.~N. Carenza, G.~Gonnella, A.~Lamura, G.~Negro, A.~Tiribocchi, Lattice
  boltzmann methods and active fluids, in press for European Physical Journal E
  30 (2019) 329--364.

\bibitem{brackbill1992}
J.~U. Brackbill, D.~B. Kothe, C.~Zemach, A continuum method for modeling
  surface tension, J. Comput. Phys. 100 (1992) 335--354.

\bibitem{chan2011film}
D.~Y. Chan, E.~Klaseboer, R.~Manica, Film drainage and coalescence between
  deformable drops and bubbles, Soft Matter 7~(6) (2011) 2235--2264.

\bibitem{williams1982}
M.~B. Williams, S.~H. Davis, Nonlinear theory of film rupture, J. Colloid
  Interface Sci. 90 (1982) 220--228.

\bibitem{thampi2013isotropic}
S.~P. Thampi, S.~Ansumali, R.~Adhikari, S.~Succi, Isotropic discrete laplacian
  operators from lattice hydrodynamics, Journal of Computational Physics 234
  (2013) 1--7.

\bibitem{kupershtokh2010criterion}
A.~L. Kupershtokh, Criterion of numerical instability of liquid state in lbe
  simulations, Computers \& Mathematics with Applications 59~(7) (2010)
  2236--2245.

\end{thebibliography}

\end{document}